\documentclass[14pt,aps,prl,twocolumn,nofootinbib,showpacs,superscriptaddress,groupedaddress]{revtex4-1}
\usepackage{graphicx,epsfig}
\usepackage{subfigure}
\usepackage{color}
\usepackage{amsmath}
\usepackage{bm}
\usepackage{multirow}
\usepackage{natbib}
\begin{document}

\title {The role of pair correlation function in the dynamical transition predicted by the mode coupling theory}

\author{Manoj Kumar Nandi$^1$, Atreyee Banerjee$^1$, Chandan Dasgupta$^2$, Sarika Maitra Bhattacharyya$^1$}
\email{mb.sarika@ncl.res.in}

\affiliation{$^1$ Polymer Science and Engineering Division, CSIR-National Chemical Laboratory, Pune-411008, India}

\affiliation{$^2$ Centre for Condensed Matter Theory, Department of Physics, Indian Institute of Science, Bangalore 560012, India}

\date{\today}

\begin{abstract}
In a recent study we have found that for a large number of systems the configuration entropy at pair level, $S_{c2}$, which is 
primarily determined by the structural information, vanishes at the mode coupling transition temperature $T_{c}$. 
 Thus it appears that the information of the 
transition temperature is embedded in the structure of the liquid. In order to investigate this we describe the dynamics of the 
system at the mean field level and using the concepts of the dynamical density function theory show that the dynamics depends only 
on the structure of the liquid. Thus this theory is similar in spirit to the microscopic MCT. However unlike microscopic MCT, which predicts
a very high transition temperature,
the present theory predicts a transition temperature which is similar to $T_{c}$. Thus our study reveals that the information of 
the mode coupling transition temperature is embedded in the structure of the liquid.
\end{abstract}
\maketitle

In the theory of liquid state the structure which is also directly accessible in experiments, plays an important role. 
For systems with  pair additive potentials,  
the thermodynamic quantities, are
 written in terms of the pair correlation function. One such thermodynamic quantity, the excess entropy 
 can be expanded into many body terms and its two body contribution, $S_{2}$  is given in terms
 of the pair correlation function \cite{Kirkwood,green_jcp,raveche,Wallace}.
 In a recent work by some of us we have defined another quantity,
 the pair configurational entropy $S_{c2}$ which can
 be derived from
 $S_{2}$ and the vibrational entropy, $S_{vib}$ \cite{bssb}. We have shown that both $S_{2}$ and $S_{c2}$ are sensitive to small 
 changes in the structure and have also shown that
 $S_{c2}$ appears to vanish at the mode coupling transition temperature, $T_{c}$\cite{bssb,manoj_unravel,ref_ntw_wahn}. 
 $T_c$ is determined by fitting the $\alpha$ relaxation time obtained from simulation studies \cite{szamel-pre,kim_saito,manoj_unravel} or 
 experiments \cite{du_mct_exp,mct_exp_prl2} to power
 law behaviour as predicted by both the standard mode coupling theory (MCT) $F_2$ 
 model \cite{Gotze} and also the schematic form of the 
 generalised MCT \cite{reichman-gmct}. 
However although the data fits the power law behaviour over a certain temperature regime, the transition predicted by MCT is avoided in these
real systems. 
Moreover, the microscopic MCT using structural information contained in the pair correlation function as obtained from simulation
studies, predicts a transition temperature, $T_c^{micro}$ which is higher than $T_c$.
Thus there has been
 debate about the physical meaning of these transition temperatures. It has been found that the p-spin model where 
 the free energy barriers for activated motion diverge in the thermodynamic limit shows a MCT
 like power law behaviour and a similar transition temperature \cite{RFOT}. 
 The concepts of the p-spin model has then been extended to real system \cite{cavagna2001_epl} where it has been shown that at $T_c$
 the saddle order in the landscape  appear to
 vanish \cite{sciortino-pre-2002,sciortino-saddle} and the activated dynamics starts to play a dominant role in the total 
 dynamics \cite{sarika_PNAS}.

In this letter we try to understand the origin of the connection between the vanishing of $S_{c2}$ and MCT transition. The fact that 
we have observed this connection for multiple
 systems also at different state points rules out the spurious nature of this connection \cite{bssb,manoj_unravel,ref_ntw_wahn}. 
Note that in the calculation of $S_{c2}$ apart from the information of the vibrational entropy the only other information that is required is that of the 
structure of the liquid through its pair correlation
 function.  Surprisingly as mention before identical information of the pair interaction, given by the static structure factor, when 
 fed into standard microscopic MCT, predicts a transition
 temperature which is much higher than $T_{c}$ \cite{Reichman,szamel-pre}. The transition predicted by the microscopic MCT is 
 attributed to the non-linear feedback
 mechanism present in the theory and to the temperature dependence of the structure factor. Thus the transition is a coupled effect 
 which, within the MCT, cannot be separated. 
Hence, two independent formalism using the same information predicts two different transition temperatures. 

In this work we attempt to verify if indeed there is any information of transition temperature embedded in the pair correlation function and if so which is that temperature.
 Towards this goal we propose a completely 
different theoretical framework to ensure that we are able to avoid any artifact of the approximations made in the earlier two 
theories.

MCT can be viewed as a mean-field description because it becomes exact
for certain systems with infinite-range interactions. In analogy with
mean-field descriptions of spin systems in which the thermodynamics of
a system of interacting spins is approximated by that of a single spin
in an effective field, we formulate a description of the dynamics of a
collection of interacting particles in terms of that of a single
particle in an effective potential which may be viewed as the ``caging
potential'' created by the neighbours of the particle being
considered. Using arguments similar to those 
as presented earlier \cite{wolynes,Schweizer}, we obtain
an effective potential in terms of the equilibrium pair correlation
function and calculate the mean first passage time for activated
escape from this potential. We find that the temperature dependence of
the inverse of this time scale follows that of the diffusivity
obtained from simulations for several model glass-forming liquids. In
particular, the temperature dependence of the inverse of the time
scale exhibits approximate power-law behaviour that extrapolates to
zero at a temperature that is very close to the dynamic transition
temperature of MCT obtained by fitting the temperature dependence of
the diffusivity to a power law. These results show that the pair
correlation function contains all the information needed to determine
the MCT transition temperature.


We consider an over-damped dynamics of N interacting particles. The evolution of the N-body distribution function
$P_N({\bf{r_1,...,r_N}},t)$ is given by the Fokker-Planck equation \cite{fokker},
\begin{equation}
 \frac{\partial P_N}{\partial t}=\sum_{i=1}^N\nabla_i.\left[\frac{k_BT}{\xi}\nabla_i P_N
 + \frac{P_N}{m\xi} \nabla_i U({\bf{r_1}},....,{\bf{r_N}})\right],
 \label{fokker-planck}
\end{equation}
where $P_N({\bf{r_1}},....,{\bf{r_N}})$ is the N-body distribution function,
$U({\bf{r_1,...,r_N}},t)=\sum_{i<j}u(|{\bf{r_i-r_j}}|)$ is the interaction potential of the system. 
 $\xi$ and m are the friction of the system and mass of a particle respectively.
 For simplicity we choose mass m=1.

If we write the Eq.(\ref{fokker-planck}) for reduced probability distribution 
$P_j({\bf{r_1,...,r_j}},t)=\int P_N({\bf{r_1,..,r_N}},t)d{\bf{r}}_{j+1}....d{\bf{r}}_N$, we will get a BBGKY like hierarchy.
The first equation of this hierarchy is,
\begin{equation}
 \frac{\partial P_1}{\partial t}=\nabla_1.\left[\frac{k_BT}{\xi}\nabla_1 P_1+\frac{(N-1)}{\xi}
 \int P_2 \nabla_1 u_{12} d{\bf{r_2}}\right].
 \label{bbgky1}
\end{equation}
We can write the above equation in terms of 
n-particle density distribution $\rho^{(n)}({\bf{r}}^n,t)$, where  
 $\rho^{(n)}({\bf{r}}^n,t)=\frac{N!}{(N-n)!}P_n({\bf{r}}^n,t).$
Thus we can write  Eq.(\ref{bbgky1}) for the evolution of averaged density
$\rho({\bf{r}},t)=\rho^{(1)}({\bf{r}},t)$.
Next we can make a mean-field approximation where the second term in Eq.(\ref{bbgky1}) can be replaced by an effective
mean-field potential $\Phi({\bf{r}})$.

\begin{equation}
\frac{\partial \rho({\bf{r}},t)}{\partial t}=\nabla.\left[\frac{1}{\xi}\Big( k_BT\nabla \rho({\bf{r}},t)+\rho({\bf{r}},t) \nabla \Phi({\bf{r}})\Big)\right].
\label{smol-eq}
\end{equation}
Where $\rho({\bf{r}},t) \nabla \Phi({\bf{r}})= \int \rho^{(2)}({\bf{r,r'}},t) \nabla u(|{\bf{r-r'}}|)d{\bf{r'}}$.
Thus in the spirit of mean-field theory Eq.(\ref{smol-eq}) can also describe the dynamics of a set of non-interacting
particles in an external potential $\Phi(r)$.

To calculate the potential we use the dynamic density-functional theory (DDFT) approach, where two body equilibrium density
$\rho^{(2)}({\bf{r,r'}})$ is connected to the gradient of one body direct correlation function, $C^{(1)}({\bf{r}})$ through the sum
rule \cite{Evans_AdvPhys}.
$C^{(1)}({\bf{r}})$ is the functional derivative of the excess free energy $F_{ex}$. The standard form for excess free energy is
\cite{ramakrishnan-yossuf},
 $F_{ex}\simeq -\frac{1}{2}\int d{\bf{R}}\int d{\bf{R'}} \rho({\bf{R}})C(|{\bf{R-R'}}|)\rho({\bf{R'}})
         = -\frac{1}{2} \int \frac{d{\bf{q}}}{(2\pi)^3}C(q)\rho^2({\bf{q}}).$
Where $\langle\rho({\bf{R}})\rangle=\langle \sum_i \delta ({\bf{R-R_i}})\rangle$ and $C(|{\bf{R-R'}}|)$ is the direct correlation function. 
We follow a set of arguments given in Ref. 20, 21 
to obtain the following form for the effective potential: (See SI)
\begin{eqnarray}
 \Phi(r)=\frac{\delta F_{ex}[\rho(r)]}{\delta \rho({\bf{r}})} \simeq -\frac{1}{2}\int \frac{d{\bf{q}}}{(2\pi)^3} \rho C^2(q) S(q) e^{-q^2r^2/3},
\end{eqnarray}
Here $S(q)$ is the static structure factor of the system.
Note that the mean field potential $\Phi(r)$ can be 
described only in terms of pair structure of the liquid. 

Eq.(\ref{smol-eq}) can be identified as a
Smoluchowski equation and by using standard formalism, we can calculate the mean first passage
time $\tau_{mfpt}$ for the system \cite{zwanzig} (See SI),
\begin{equation}
 \tau_{mfpt}=\frac{1}{D_0}\int_0^{L/2} e^{\beta \Phi(y)}dy \int_0^{L/2} e^{-\beta \Phi(z)}dz.
 \label{tau-mfpt}
\end{equation}
Where $D_0=k_BT/\xi$ and L is the simulation box length.

\begin{figure*}[ht!]
\centering
 \subfigure{
 \includegraphics[width=0.4\textwidth]{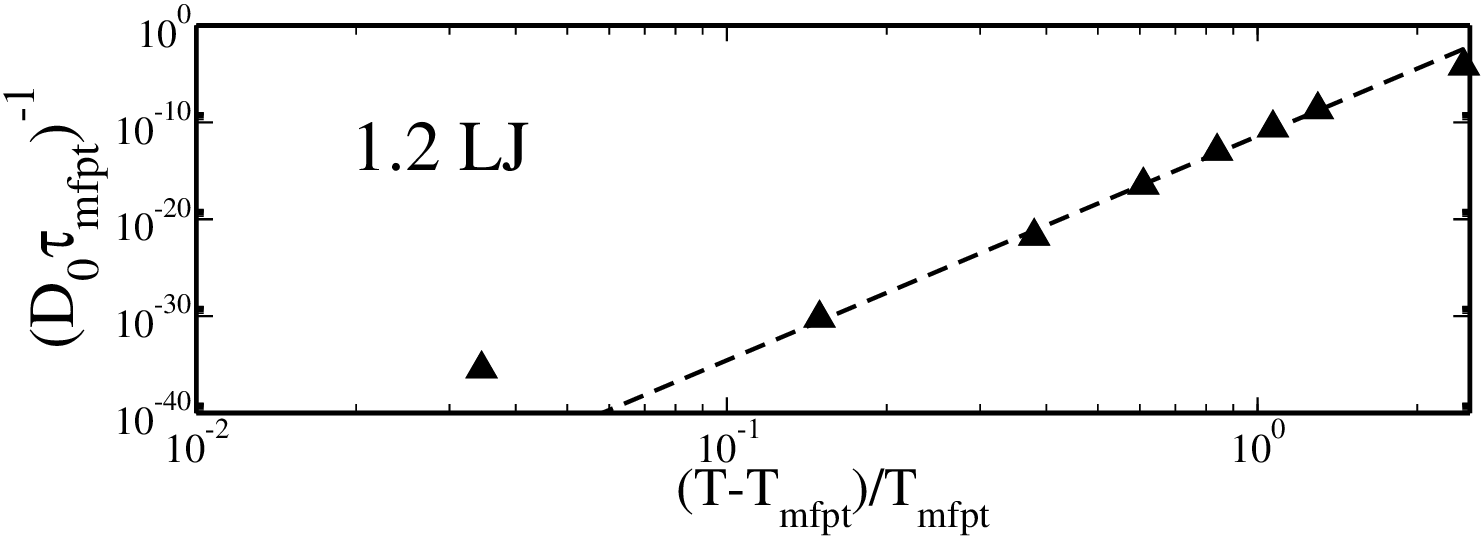}}
\subfigure{
 \includegraphics[width=0.4\textwidth]{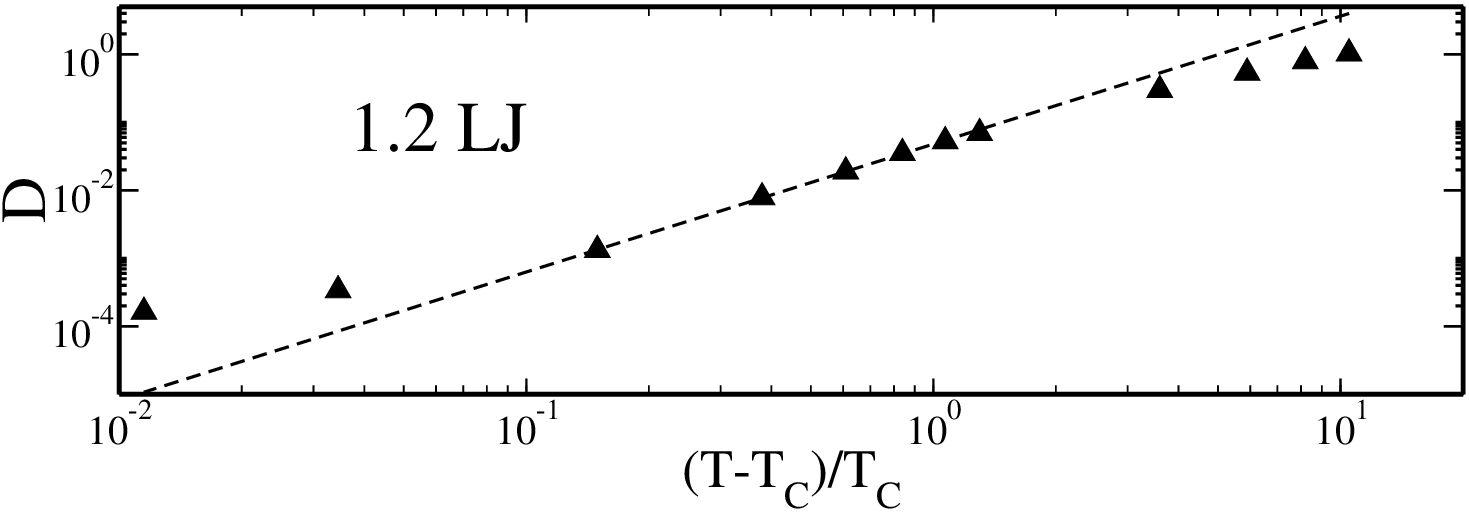}}
\subfigure{
 \includegraphics[width=0.4\textwidth]{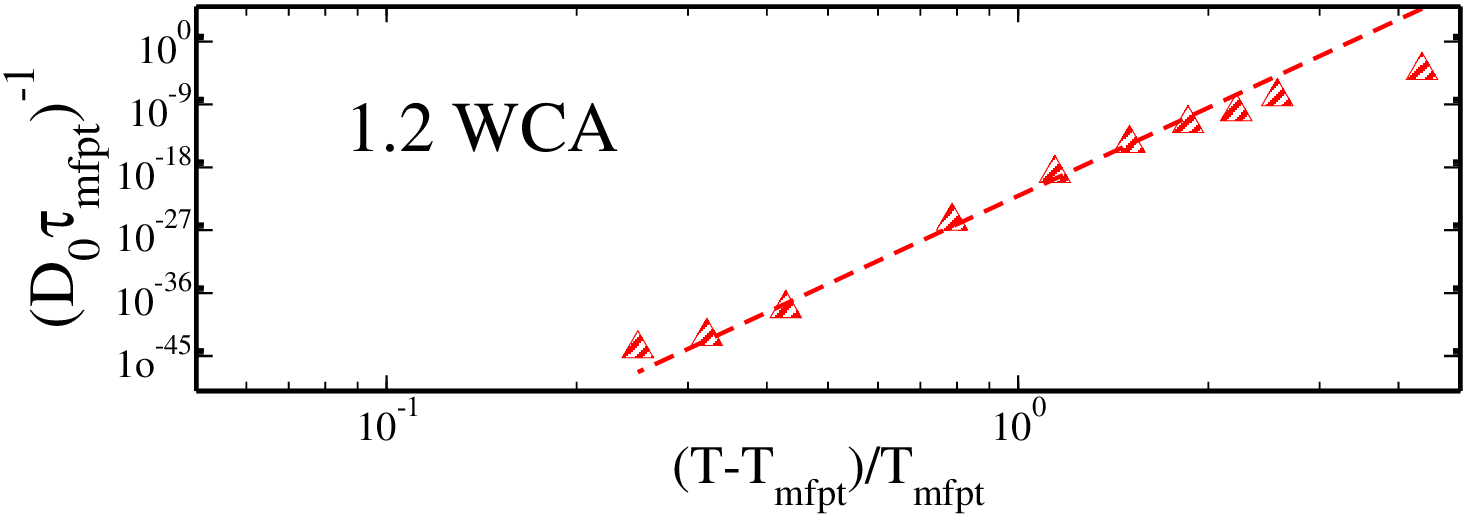}}
\subfigure{
 \includegraphics[width=0.4\textwidth]{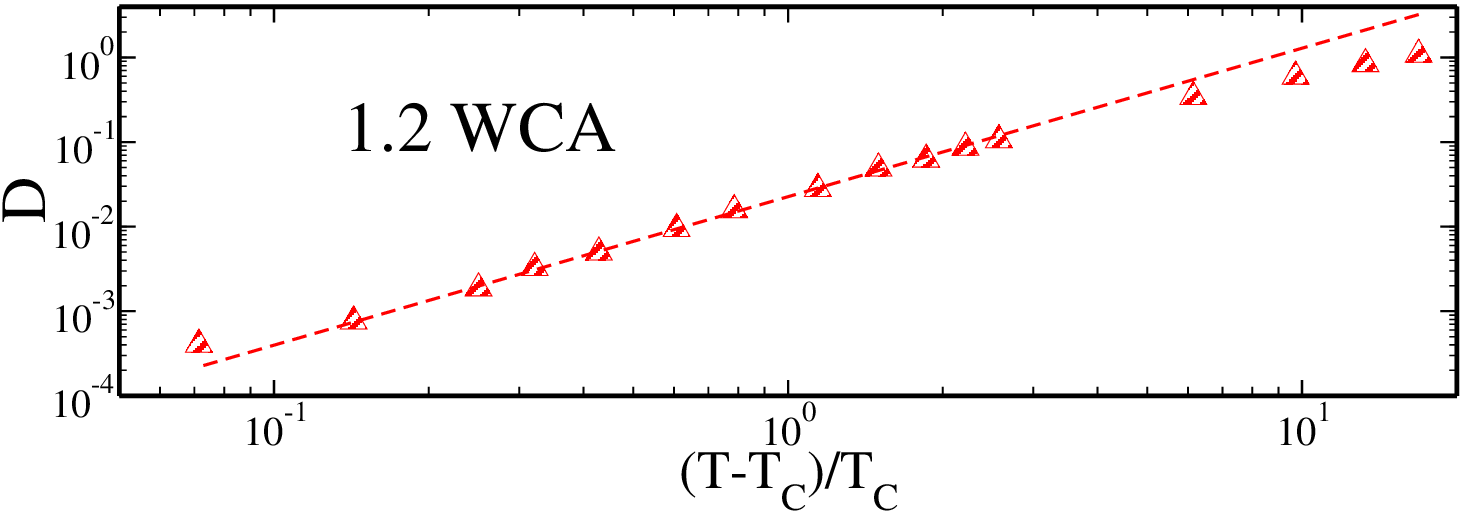}}
\subfigure{
 \includegraphics[width=0.4\textwidth]{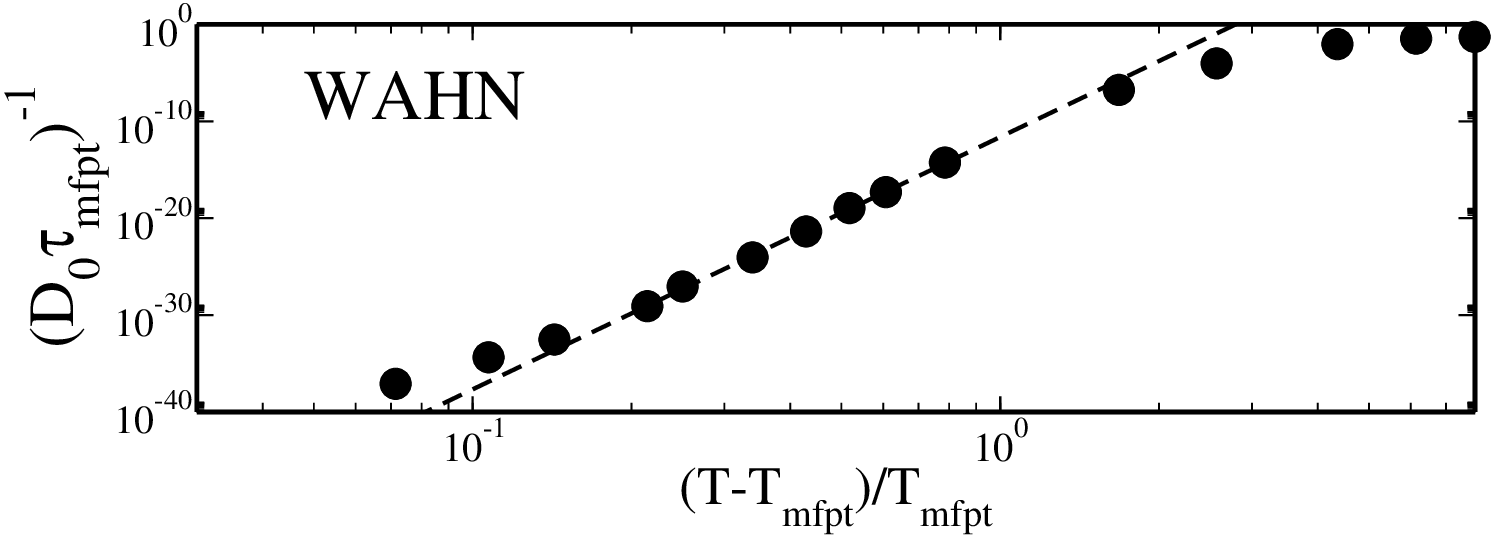}}
\subfigure{
 \includegraphics[width=0.4\textwidth]{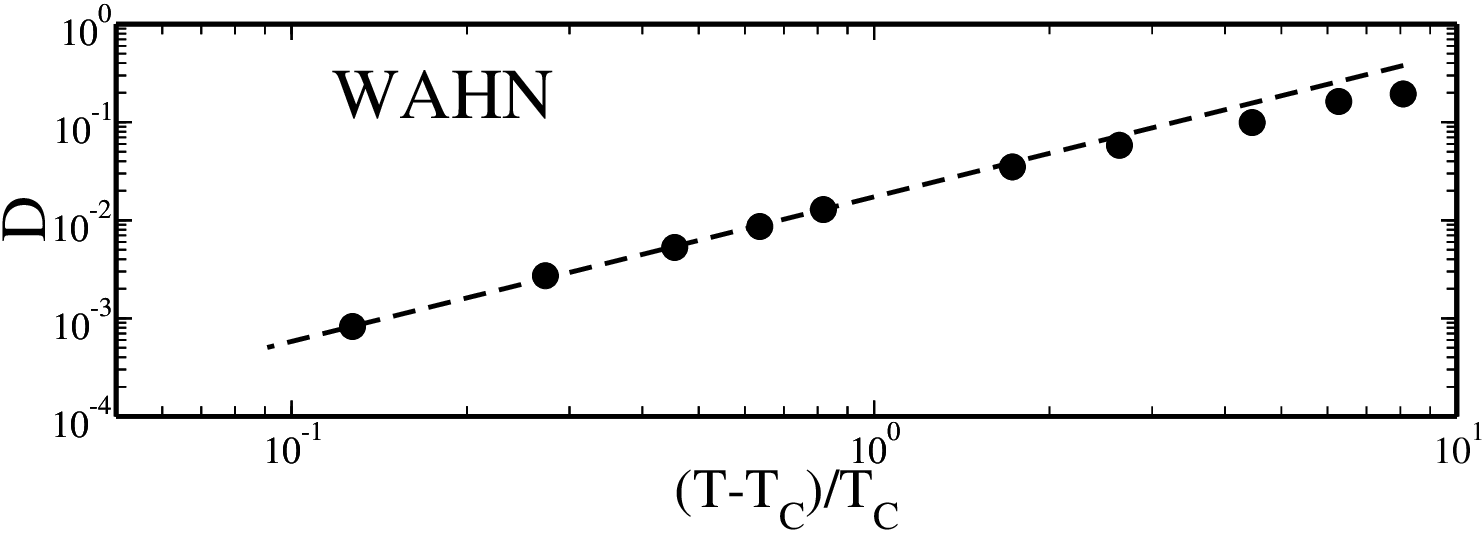}}
\subfigure{
 \includegraphics[width=0.4\textwidth]{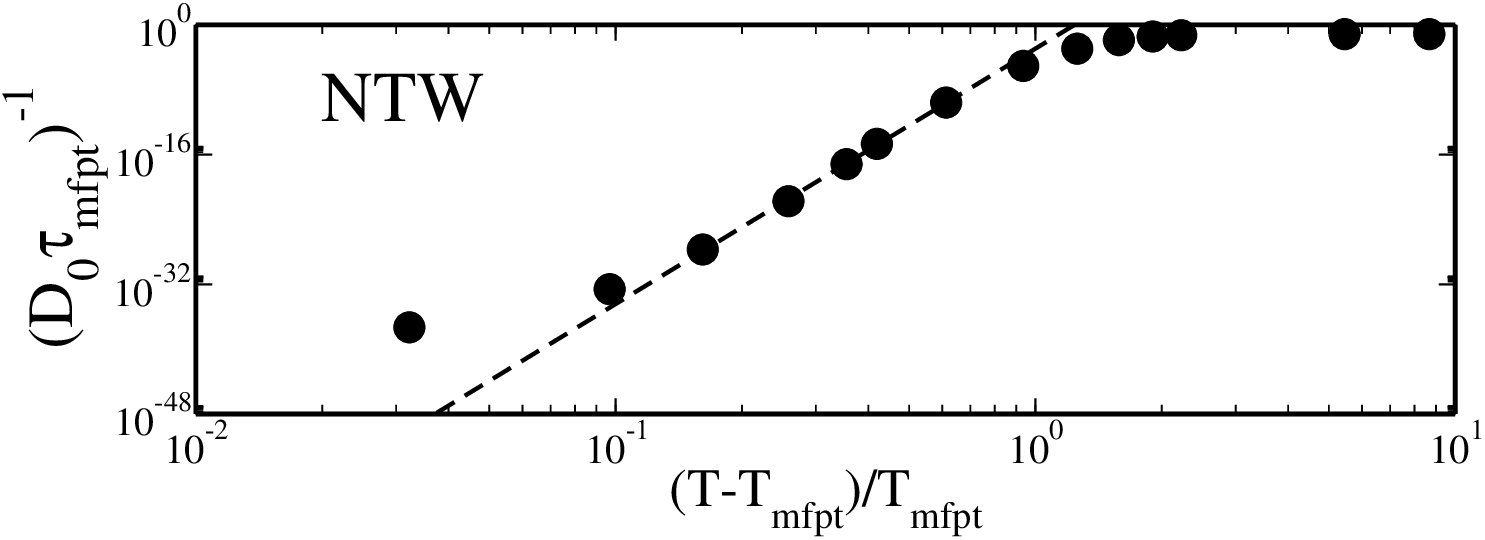}}
\subfigure{
 \includegraphics[width=0.4\textwidth]{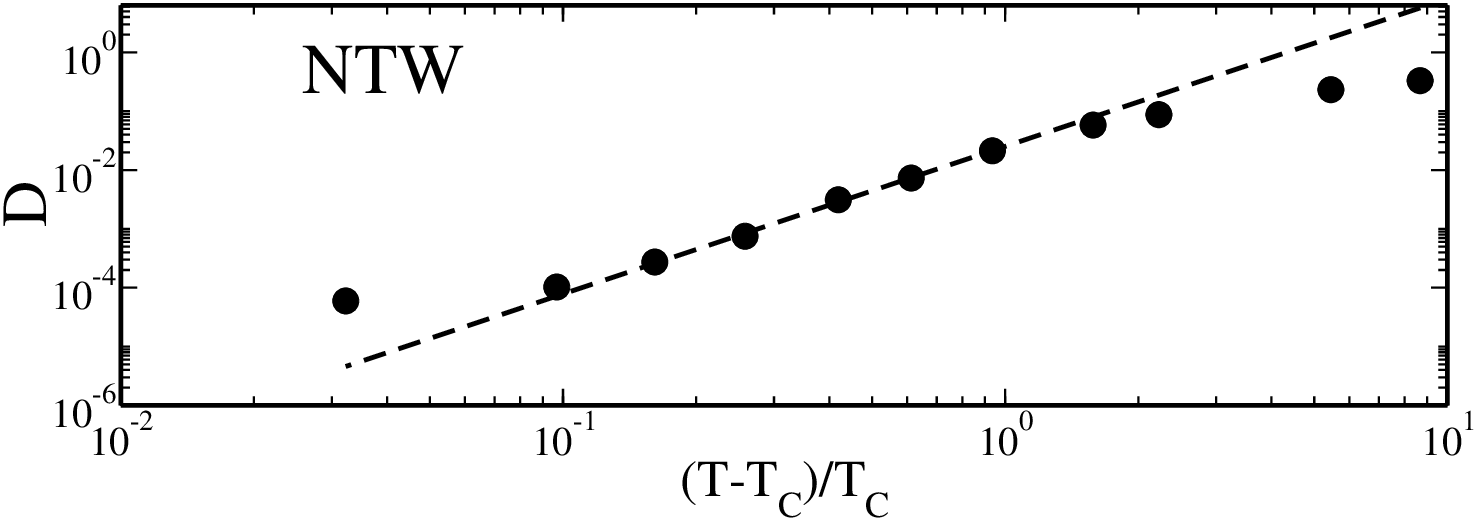}}
\subfigure{
 \includegraphics[width=0.4\textwidth]{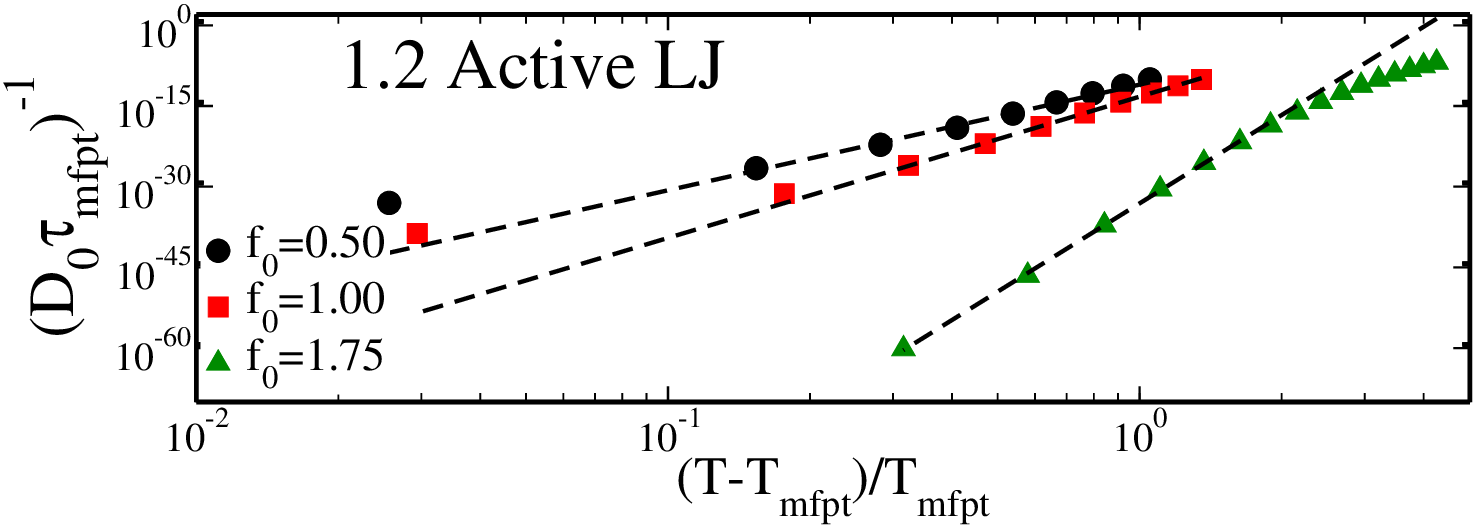}}
\subfigure{
 \includegraphics[width=0.4\textwidth]{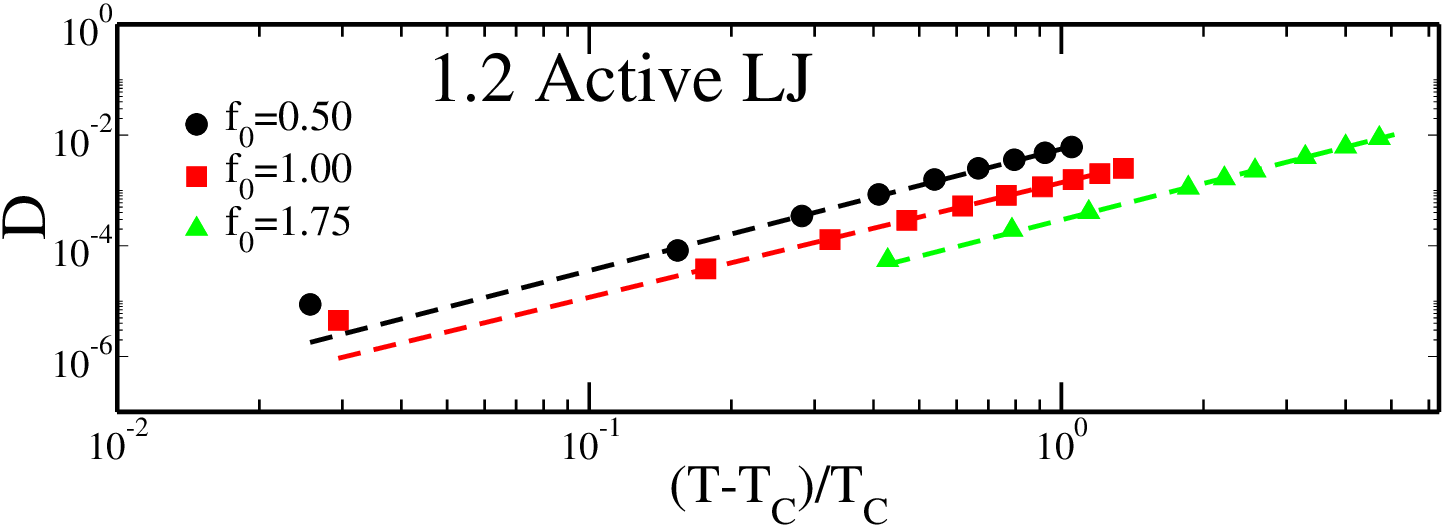}}
\caption{ \it{ The power law dependence of $(D_0\tau_{mfpt})^{-1}$  predicts a transition temperature $T_{mfpt}$ (left panel)
and the same for total diffusivity (right panel). The dashed lines are the power law fit. For active systems $f_0$ is the activity
as described in Ref. 28. 
For clarity $ln D$ plot is shifted for $f_0 = 1.0$ by -1. Here for all the systems we take $D_0=1$.
}}
\label{s_plot_ns_D}
\end{figure*}

In Fig.\ref{s_plot_ns_D} we show some representative plots of the temperature dependence of the inverse mean first passage time 
for different model systems \cite{system_details_new}. 
We also plot the corresponding diffusion (D) values for these systems. We find that for each 
of these systems the $1/D_0\tau_{mfpt}$ shows a power law
divergence and can predict a transition temperature ($T_{mfpt}$) of the respective systems given in Table \ref{table1}. 
  Note that not only the temperature $T_{mfpt}$, is similar to the critical temperature
obtained from the MCT power law behaviour of diffusion, $T_c$ (also given in Table \ref{table1}), but also the temperature regime 
over which $1/D_0\tau_{mfpt}$ shows linearity 
is similar to that of the diffusion (Fig.\ref{s_plot_ns_D})/ relaxation time \cite{manoj_unravel}.

 \begin{figure}[ht!]
\centering
\subfigure{
 \includegraphics[width=0.35\textwidth]{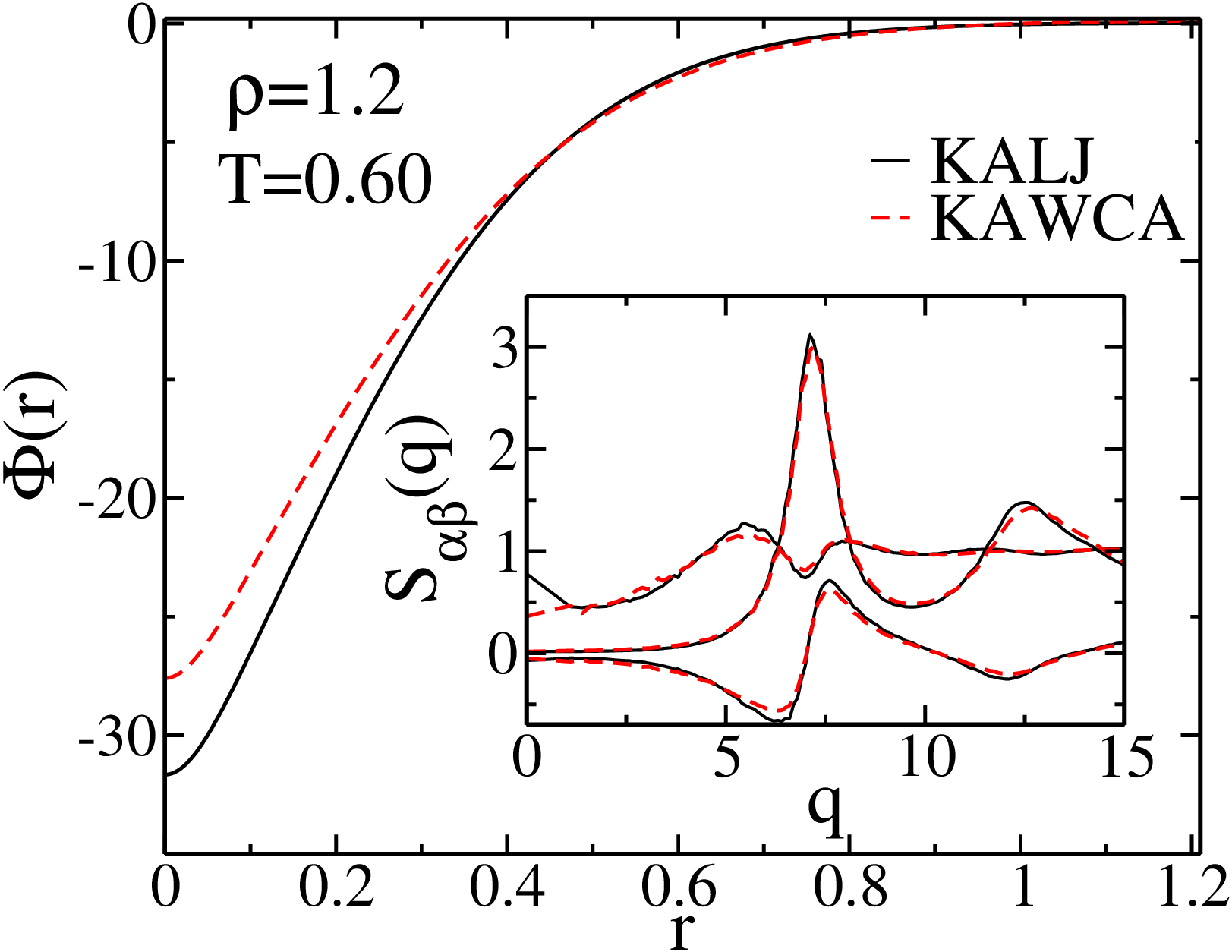}}
\caption{\it{ Plot of  potential energy surface $\Phi(r)$ vs. r at temperature T=0.60 and density $\rho=1.2$ for KALJ and KAWCA systems.
(Inset) The plots for partial structure factor $S_{\alpha\beta}(q)$ vs. q.
}}
\label{fig-potential}
\end{figure}
\begin{table*}[ht!]
\begin{center}
\caption{\it{Transition temperatures ($T^*$) are shown. $T_{mfpt}$ is obtained from the power law fits of $1/D_0\tau_{mfpt}$ with 
temperature. $T_c$ values are either obtained by fitting the diffusion values to power law or from earlier reported studies,
KALJ\cite{tarjus_pre}, KAWCA \cite{tarjus_pre}, active LJ \cite{active-data}, WAHN \cite{wahnstrom} and NTW \cite{ntw-biroli}.   
 $T_c^{micro}$ is calculated from solving the microscopic MCT equation 
 \cite{szamel-pre,tarjus_pre,manoj_unravel} (Eq.(\ref{mct2})). }}
\small
\begin{tabular}{|c|c|c|c|c|c|c|c|c|c|c|c|c|}
\hline
\multirow{2}{*}{$T^*$} & \multicolumn{3}{c|}{KALJ} & \multicolumn{3}{c|}{KAWCA} & \multicolumn{3}{c|}{Active LJ}& \multirow{2}{*}{NTW} & %
    \multirow{2}{*}{WAHN} \\
\cline{2-10}
 & $\rho=$1.2 &$\rho=$ 1.4 & $\rho=$1.6 & $\rho=$1.2 & $\rho=$1.4 & $\rho=$1.6 & $f_0=0.50$ & $f_0=1.00$& $f_0=1.75$ & & \\
\hline
 $T_{mfpt}$&0.428 & 0.94 & 1.757 & 0.283&0.824 & 1.691& 0.38& 0.335 & 0.196& 0.308 & 0.566\\
   &$\pm$0.022 & $\pm$0.029 & $\pm$0.042 & $\pm$0.005& $\pm$0.04 & $\pm$0.018& $\pm$0.004& $\pm$0.006 & $\pm$0.013& $\pm$0.012&$\pm$0.013\\
\hline
 $T_c$ &0.435 &0.93 &1.76 &0.28 &0.81 &1.69 &0.39 &0.34 &0.19 &0.31 &0.56 \\
\hline
 $T_c^{micro}$ &0.887 &1.868 &3.528 & 0.76&1.771 & 3.33& 0.768 & 0.761 & 0.747 &0.464 &0.87\\
\hline
\end{tabular}
\label{table1}
\end{center}
\end{table*}
 To show that the present analysis is sensitive to small changes in structure, we compare the results of KALJ and KAWCA systems
at $\rho=1.2$.
In Fig.\ref{fig-potential} we plot $\Phi(r)$ which is used to calculate $\tau_{mfpt}$ (via Eq.(\ref{tau-mfpt})) and show that 
although the structures are similar, the one body potentials are different. This difference is enough to predict the difference 
in $T_{mfpt}$ (see Table \ref{table1}). This observation is similar 
to that presented in an earlier study, where we have shown that although the radial distribution functions are quite 
similar, the pair configurational entropy ($S_{c2}$) for these systems are different \cite{atreyee-Sc2}. Thus the temperature at which
the $S_{c2}$ vanishes are also different.

\begin{figure}[ht!]
\centering
\includegraphics[width=0.35\textwidth]{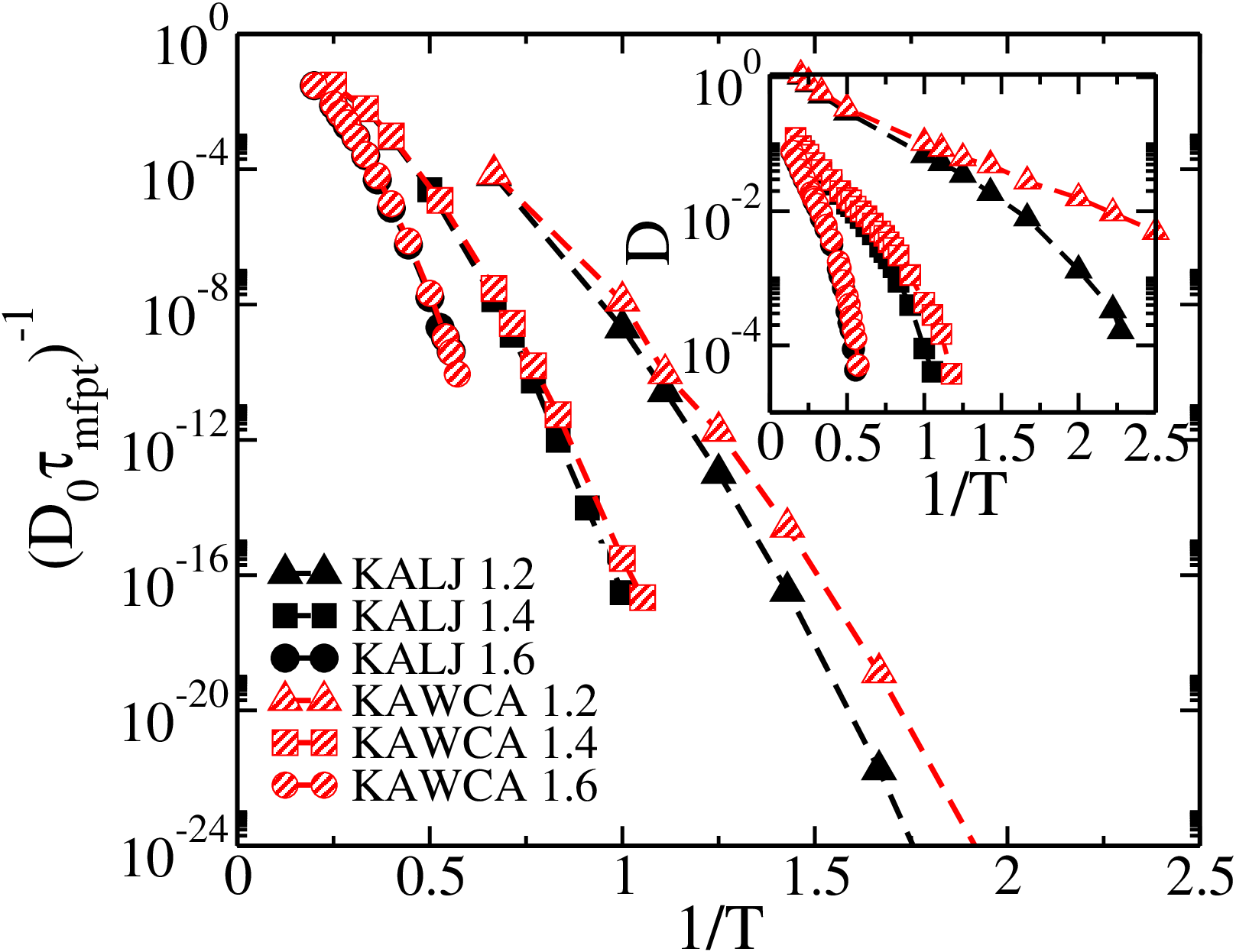}
\caption{\it{ The plot of $(D_0\tau_{mfpt})^{-1}$ against inverse temperature for KALJ and KAWCA systems at different densities. 
We take $D_0=1$. 
(Inset) The plot for diffusion coefficients with inverse temperature. }}
\label{densities-dep-plot}
\end{figure}

Further we show that this method of mean first passage time calculation can also predict the density effect. In Fig.\ref{densities-dep-plot} 
we plot the $1/D_0\tau_{mfpt}$ for KALJ and KAWCA systems for three different
 densities.  We also plot the following diffusion coefficients in the inset. Note that the $1/D_0\tau_{mfpt}$ show similar behaviour
 as the diffusion coefficients.
At low  density the values for the two systems are apart and at high density they overlap. 


Next we show that we can derive the standard microscopic MCT equation from our present formalism.
 Eq.(\ref{smol-eq}) when expressed in terms of the fluctuation of density, $\delta \rho$
 ($\delta \rho(r)=\rho(r)-\rho_0$, where $\rho_0$ is the averaged
 density of the system) and $\Phi(r)$ is replaced in terms of $F_{ex}$, can be written as,
\begin{eqnarray}
\xi\frac{\partial \delta \rho({\bf{r}},t)}{\partial t}=\nabla.[k_BT\nabla \delta \rho({\bf{r}},t)\nonumber\\
               + (\delta \rho({\bf{r}},t)+\rho_0) \nabla \frac{\delta F_{ex}( \rho({\bf{r}}))}{\delta \rho({\bf{r}})} ].
\label{eq10}
\end{eqnarray}
Next in Eq.(\ref{eq10}) we use the expression\cite{ramakrishnan-yossuf} of $F_{ex}$.
After taking the functional derivative of excess free energy and then the gradient of it,
we can write Eq.(\ref{eq10}) as,
\begin{eqnarray}
 \xi\frac{\partial \delta \rho_{\bf{k}}(t)}{\partial t}=-{k_BTk^2}(1-\rho_0C_k)\delta \rho_{\bf{k}}(t)\nonumber\\
               +\frac{k_BT}{2}\int_q{\bf { dq}}[{\bf{k.q}}C_q+{\bf{k.(k-q)}}C_{|\bf{k-q}|}]\delta \rho_{\bf{q}}(t)\delta \rho_{{\bf{k-q}}}(t).
\label{mct}
\end{eqnarray}
 Eq.(\ref{mct}) is the Langevin equation for the density field. Following Kawasaki's arguments, this equation can be converted
 into standard MCT equation \cite{kawasaki,snandi}. The last non-linear term on the right hand side can be identified 
 as the fluctuating force. In non-Markovian limit applying fluctuation dissipation relation (FDR) \cite{kawasaki},
 the friction $\xi$ can be replaced by short time part of friction $\gamma$ and 
 a memory function $\mathcal{M}(k,t)$. 
 Thus we can rewrite Eq.(\ref{mct}) as,
 \begin{eqnarray}
 \gamma \frac{\partial \delta \rho_{\bf{k}}(t)}{\partial t}+ \frac{k_BTk^2}{S(k)}\delta \rho_{\bf{k}}(t)\nonumber\\
 +\int_0^t ds\mathcal{M}(k,t-s)\frac{\partial \delta \rho_{\bf{k}}(t)}{\partial s}+\mathcal{R}_{\bf{k}}(t)=0,
  \label{langevin-1}
 \end{eqnarray}
 where $\mathcal{R}_{\bf{k}}(t)$ is the new thermal noise and 
 $\mathcal{M}(k,t)=\frac{\langle\mathcal{R}_k(t)\mathcal{R}_{-k}(0)\rangle}{k_BTV} =\frac{k_BT\rho_0}{16\pi^3}\int d{\bf{q}}
\hat [k.({\bf{q}}C_q+{\bf{(k-q)}}C_{|\bf{k-q}|})]^2S_q(t)S_{k-q}(t)$ (see SI).
Here we use Gaussian decoupling and Wick's theorem to treat the four-point 
 correlation function.
 We can write a mode coupling theory equation in the over-damped limit for density-density 
 correlation $S_k(t)=\langle\delta \rho_k(t)\delta \rho_{-k}(0)\rangle$. 
 \begin{equation}
 \gamma\frac{\partial S_k(t)}{\partial t}+\frac{k_BTk^2}{S_k}S_k(t)+\int \mathcal{M}(t-\tau)\dot S_k(t) d\tau=0.
\label{mct2}
\end{equation}
In Table \ref{table1} we present the transition temperatures predicted by Eq.(\ref{mct2}). We find that for all the systems 
$T_c^{micro} >> T_c$. This higher value of $T_c^{micro}$ can be connected to the approximations used to arrive from Eq.(\ref{smol-eq}) to
Eq.(\ref{mct2}).
In a recent work we have also shown that the form of the vertex function in the theory which depends on the structure factor might 
also be responsible for this premature divergence \cite{manoj_unravel}.

  In the present work in analogy with mean-field descriptions of spin systems, we reduce the dynamics of the 
N-interacting particle system given by Fokker-Planck equation to that of N-non-interacting particle system in an effective
external potential
where the latter provides the effect of the interaction only at the two body level. Thus in our formalism by construction we do not 
have any effect of correlation beyond two body. 
We then obtain the mean first passage time, which 
 now depends only on the structure of the liquid.
The temperature dependence of it, can predict a 
transition temperature which is similar to $T_c$. Thus we conclude that the information of the MCT transition temperature is 
embedded in the pair structure 
of the liquid and it is the mean field transition temperature. 
In an earlier work, we have shown that the pair configurational entropy ($S_{c2}$) vanishes at $T_c$ \cite{atreyee-Sc2}. At infinite
dimension $S_{c2}$ should be the total configurational entropy and the temperature at which it vanishes should be the Kauzmann 
temperature. Also for those systems the Kauzmann temperature and the mean field transition temperature should be similar. Thus
our earlier result of $S_{c2}$ vanishing at $T_c$ is consistent and similar to the present finding.
In this present work we further
show that starting from the same Fokker-Planck equation we can also derive the microscopic MCT equation. However the transition
temperature predicted by the microscopic MCT is much higher than the $T_c$ value.
Thus the break down of microscopic MCT in predicting the transition temperature, can be connected to the Gaussian decoupling  
approximation. However our present formalism although predicts the correct transition temperature, unlike the microscopic MCT 
 the power law exponent is not universal. 
 Also note that this present method of deriving the transition temperature is intimately connected to the structure of the liquid.
Thus for systems 
like pinned particles, as the structure remains same, this theory will not be able to predict different transition temperatures
for different pinning densities.

\begin{thebibliography}{10}
\bibitem{ref_ntw_wahn}
The $T_{K2}$ values for WAHN and NTW model have been found to be similar with
  $T_c$ which will be reported in future communication .
\end{thebibliography}

\end{document}


\title{The role of pair correlation function in the dynamical transition predicted by the mode coupling theory-Supplementary Information}
\author{Manoj Kumar Nandi}
\affiliation{\textit{Polymer Science and Engineering Division, CSIR-National Chemical Laboratory, Pune-411008, India}}

\author{Atreyee Banerjee}
\affiliation{\textit{Polymer Science and Engineering Division, CSIR-National Chemical Laboratory, Pune-411008, India}}

\author{Chandan Dasgupta}
\affiliation{\textit{Centre for Condensed Matter Theory, Department of Physics, Indian Institute of Science, Bangalore 560012, India}}

\author{Sarika Maitra Bhattacharyya}
\email{mb.sarika@ncl.res.in}
\affiliation{\textit{Polymer Science and Engineering Division, CSIR-National Chemical Laboratory, Pune-411008, India}}

\maketitle

\section{Details of the derivation of Eq.(3)}
We start from the N-body Fokker-Planck equation for interacting particles.
\begin{equation}
 \frac{\partial P_N}{\partial t}=\sum_{i=1}^N\nabla_i.[\frac{k_BT}{\xi}\nabla_i P_N
 + \frac{P_N}{m\xi} \nabla_i U({\bf{r_1}},....,{\bf{r_N}})]
 \label{eq1}
\end{equation}
Where, $P_N({\bf{r_1}},....,{\bf{r_N}},t)$ is the N-body probability distribution function,
$U({\bf{r_1}},....,{\bf{r_N}})=\sum_{i<j}u_{ij}$, for simplicity we use m=1.
Now we can write Eq.(\ref{eq1}) for reduced probability density
$P_j({\bf{r_1}},...,{\bf{r_j}},t)=\int P_N({\bf{r_1}},....,{\bf{r_N}},t)d{\bf{r_{j+1}}}...d{\bf{r_N}}$ as,

\[
 \frac{\partial P_j}{\partial t} =\sum_{i=1}^j\nabla_i.[\frac{k_BT}{\xi}\nabla_i P_j
 + \frac{P_j}{\xi} \sum_{k=1}^j\nabla_i u_{i,k}
 \]
 \[
 +  \frac{N-j}{\xi} \int P_{j+1}\nabla_i u_{i,j+1}d{\bf{r_{j+1}}} ].  
\]
This equation is equivalent to BBGKY hierarchy.
The first equation of this hierarchy is,
\begin{equation}
 \frac{\partial P_1}{\partial t}=\nabla_1.\left[\frac{k_BT}{\xi}\nabla_1 P_1+\frac{(N-1)}{\xi}
 \int P_2\nabla_1 u_{12} d{\bf{r_2}}\right].
 \label{eq4}
\end{equation}
Here the second term becomes zero as we use pair potential.
We know that,
\begin{equation}
 \rho^{(n)}({\bf{r}}^n,t)=\frac{N!}{(N-n)!Z_N}P_n({\bf{r}}^n,t).
 \label{eq5}
\end{equation}
From Eq.(\ref{eq5}), we can write,
\begin{equation}
 \rho^{(2)}({\bf{r_1,r_2}},t)=N(N-1)P_2({\bf{r_1,r_2}},t).
 \label{eq6}
\end{equation}
Now with the help of Eq.(\ref{eq6}), we can write Eq.(\ref{eq4}) as,
\begin{eqnarray}
  &&\frac{\partial \rho({\bf{r_1}},t)}{\partial t}=\nonumber\\
  &\nabla_1&.\left[\frac{k_BT}{\xi}\nabla_1 \rho({\bf{r_1}},t)+
 \frac{1}{\xi}\int \rho^{(2)}({\bf{r_1,r_2}},t) \nabla_1 u_{12} d{\bf{r_2}}\right].\nonumber
\end{eqnarray}
We can now drop the particle index and write the above equation as,
\begin{equation}
\frac{\partial \rho({\bf{r}},t)}{\partial t}=\nabla.\left[\frac{1}{\xi}\Big(k_BT\nabla \rho({\bf{r}},t)+\rho({\bf{r}},t) \nabla \Phi({\bf{r}})\Big)\right], 
\label{eq7}
\end{equation}
where the form of the mean field potential $\Phi(r)$ is,
\begin{equation}
 \rho({\bf{r}},t) \nabla \Phi({\bf{r}})= \int \rho^{(2)}({\bf{r,r'}},t) \nabla u(|{\bf{r-r'}}|)d{\bf{r'}}.
 \label{eq8}
\end{equation}

To calculate the potential we use the dynamic density functional theory (DDFT) approach, where two body equilibrium density
$\rho^{(2)}({\bf{r,r'}})$ is connected to the gradient of one body direct correlation function, $C^{(1)}({\bf{r}})$ through the sum rule 
\cite{Evans_AdvPhys},
\begin{eqnarray}
 \int d{\bf{r'}}\rho^{(2)}({\bf{r,r'}})\nabla u(|{\bf{r-r'}}|)=-k_BT\rho({\bf{r}}) \nabla C^{(1)}({\bf{r}})\nonumber \\
                           =\rho({\bf{r}}) \nabla \frac{\delta F_{ex}[\rho]}{\delta \rho({\bf{r}})}.
\label{Cr_Fex}
\end{eqnarray}
The second equality follows because the one-body direct correlation is the functional derivative of the excess free energy
$F_{ex}$. The standard form for excess free energy is given by \cite{ramakrishnan-yossuf},
\begin{eqnarray}
 F_{ex}\simeq -\frac{1}{2}\int d{\bf{R}}\int d{\bf{R'}} \rho({\bf{R}})C(|{\bf{R-R'}}|)\rho({\bf{R'}})\nonumber \\
         = -\frac{1}{2} \int \frac{d{\bf{q}}}{(2\pi)^3}C(q)\rho^2({\bf{q}}),
\label{excessF}
\end{eqnarray}
where $\langle\rho({\bf{R}})\rangle=\langle\sum_i \delta ({\bf{R-R_i}})\rangle$ and $C(|{\bf{R-R'}}|)$ is the direct correlation function. Next we follow a protocol  similar to 
that presented earlier by Kenneth S. Schweizer \cite{Schweizer} and Krikpatrick {\it{et. al}} \cite{wolynes}. First we ignore
the self term in Eq.(\ref{excessF}) as we are interested in theffective interaction from all the other (N-1) particles. Next to 
dynamically close the theory at the mean field 
level, we make Vineyard approximation and write
$\rho(R,t)\simeq (3/2\pi r_i^2(t))^{3/2}exp(-3R^2/2r_i^2(t)) $. Next we drop the particle index and assume
$r_i^2(t)=r^2$, then $\rho(R)\simeq (\alpha^2/\pi)^{3/2}exp(-R^2\alpha)$, where $\alpha=3/2r^2$. So $\rho(q)=exp(-q^2/4\alpha)$
and also we ignore the i=j term, because this self term do not contribute to the force.
\begin{eqnarray}
 F_{ex}(\alpha) \simeq -\frac{1}{2}\int \frac{d{\bf{q}}}{(2\pi)^3} C(q) \{ N^{-1}\sum_{i \ne j}e^{-i{\bf{q.R^{(0)}_{ij}}}} \}
             e^{-q^2/2\alpha}. \nonumber\\.
\label{eq9si}
\end{eqnarray}
From the above equation we can calculate the potential. The form of the potential becomes,
\begin{eqnarray}
 \Phi(\alpha)=\frac{\delta F_{ex}[\rho({\bf{R}})]}{\delta \rho({\bf{R}})}
 \simeq -\frac{1}{2}\int \frac{d{\bf{q}}}{(2\pi)^3} \rho C^2(q) S(q) e^{-q^2/2\alpha},\nonumber\\
\label{potential-form}
\end{eqnarray}
where we approximate $\{ N^{-1}\sum_{i \ne j}e^{-i{\bf{q.R^{(0)}_{ij}}}} \}\simeq \rho h(q)$ \cite{wolynes} and use the 
Ornstein-Zernike relation $h(q)=C(q)S(q)$ \cite{ornstein-zernike}. Here $S(q)$ is the static structure factor of the system.\\
From Eq.(\ref{potential-form}) we see that the mean field potential only depends on r. So we can rewrite Eq.(\ref{eq7}) as,
\[
 \frac{\partial \rho(r)}{\partial t}=\frac{k_BT}{\xi}\frac{\partial}{\partial r} e^{-\Phi(r)/k_BT} \frac{\partial}{\partial r} e^{\Phi(r)/k_BT} \rho(r).
\]
\[
 \frac{\partial \rho(r)}{\partial t}=\mathcal{L}\rho(r).
\]
where $\mathcal{L}=\frac{k_BT}{\xi}\frac{\partial}{\partial r} e^{-\Phi(r)/k_BT} \frac{\partial}{\partial r} e^{\Phi(r)/k_BT}$.
This above equation can be identified as mean field Smoluchowski equation and by using standard formalism, we can calculate the mean first passage
time $\tau_{mfpt}$ for the system \cite{zwanzig},
\[
 \mathcal{L}^{\dagger}\tau_{mfpt}=-1.
\]
\[
 D_0 e^{\Phi(r)/k_BT} \frac{\partial }{\partial r}  e^{-\Phi(r)/k_BT} \frac{\partial }{\partial r} \tau_{mfpt}=-1.
\]

\begin{equation}
 \tau_{mfpt}=\frac{1}{D_0}\int_0^{L/2} e^{\beta \Phi(y)}dy \int_0^{L/2} e^{-\beta \Phi(z)}dz.
 \label{tau-mfpt}
\end{equation}
Where $D_0=k_BT/\xi$ and L is the simulation box length.
\section{Potential $\Phi(r)$ for binary syatem}
 For binary system the excess free energy becomes,
 \begin{eqnarray}
  F_{exc}\simeq -\frac{1}{2}\int d{\bf{q}}\sum_{\mu \nu}\rho_{\mu}(q)C_{\mu \nu}(q)\rho_{\nu}(q).
 \end{eqnarray}
 Using the same set of approximations used to derive Eq.(\ref{eq9si}) (SI), we can rewrite the above equation as,
 \begin{eqnarray}
  F_{exc}\simeq -\frac{1}{2}\int d{\bf{q}}\sum_{\mu \nu}C_{\mu \nu}(q)\nonumber\\
  \{\frac{1}{N}\sum_{i\ne j}^{N_\mu}\sum^{N_\nu}e^{-i{\bf{q.(R_i^\mu(0)-R_j^\nu(0))}}}\}e^{-q^2/2\alpha}.
  \label{exc_F_binary}
 \end{eqnarray}
Where $N_\mu$ and $N_\nu$ are the number of $\mu $ and $\nu$ type particles. Here we assume that for both type of particles
$\alpha$ is same. The term 
inside the bracket in Eq.(\ref{exc_F_binary}) can be approximated as $\rho x_\mu x_\nu h_{\mu \nu}(q)$, where $x_\mu$ is the 
mole fraction of component $\mu$ in the mixture and $h_{\mu \nu}(q)$ can be calculated from simulated partial static structures
$\rho x_{\mu}x_{\nu}h_{\mu\nu}(q)=S_{\mu\nu}(q)-x_{\mu}\delta_{\mu\nu}$.
Thus the potential $\Phi(\alpha)$ for binary system becomes,
\begin{eqnarray}
 \Phi(\alpha)\simeq-\frac{1}{2}\int d{\bf{q}}\sum_{\mu \nu}C_{\mu \nu}(q)\rho x_{\mu}x_{\nu}h_{\mu \nu}(q)e^{-q^2/2\alpha}.\nonumber\\
\end{eqnarray}

\section{Derivation of MCT equation from Smoluchowski equation}
If we rewrite Eq.(\ref{eq7}) interms of excess free energy and fluctuation of density ($\delta \rho(r)= \rho(r)-\rho_0$), it becomes
\begin{equation}
\frac{\partial \delta \rho(r,t)}{\partial t}=\nabla.\left[\frac{1}{\xi}\Big(k_BT\nabla \delta \rho(r,t)+(\rho_0+\delta\rho(r,t))\nabla (\frac{\delta F}{\delta \rho})\Big)\right]. 
\label{eq7exc}
\end{equation}

Now from Eq.(\ref{excessF}) after taking the functional derivative of free energy and then taking the gradient of it,
we get,
\[
 (\rho_0+\delta\rho(r,t)) \nabla (\frac{\delta \beta F}{\delta \rho})
 \]
 \[
 =-(\rho_0+\delta \rho(r,t))\nabla \int \delta \rho(r',t)C(r,r')dr', 
\]
where $C(r,r')=C(|r-r'|)$ is the two body direct correlation function.
\[
 (\rho_0+\delta\rho(r,t)) \nabla (\frac{\delta \beta F}{\delta \rho})
 \]
\[
 =-\rho_0\nabla \int \delta \rho(r',t)C(r,r')dr'  -\delta \rho(r,t)\nabla \int \delta \rho(r',t)C(r,r')dr'.
\]
\[
\nabla. \rho(r,t) \nabla (\frac{\delta \beta F}{\delta \rho})=-\rho_0\nabla^2 \int \delta \rho(r',t)C(r,r')dr'
\]
\[
                  -\nabla.\delta \rho(r,t)\nabla \int \delta \rho(r',t)C(r,r')dr'. 
\]

After taking Fourier transform, first term becomes,
\[
[-\rho_0\nabla^2 \int \delta \rho(r',t)C(r,r')dr']_k=  k^2\rho_0 C_k\delta \rho_{\bf{k}}(t),
\]
and the second term can be written as, \cite{snandi}
\[
 [\nabla.\delta \rho(x,t)\nabla \int \delta \rho(x',t)C(x-x')dx']_k=
 \]
 \[
 -\frac{1}{2}\int_{q}{\bf{dq}} [{\bf{k.q}}C_q+{\bf{k.(k-q)}}C_{|{\bf{k-q}}|}] \delta \rho_{\bf{k-q}}(t) \delta \rho_{\bf{q}}(t).
\]

So in total we get,

\begin{eqnarray}
[\nabla. \rho(r,t) \nabla (\frac{\delta F}{\delta \rho})]_k= k_BT k^2\rho_0 C_k\delta \rho_{\bf{k}}(t)\nonumber\\
+\frac{k_BT}{2}\int_q{\bf { dq}}[{\bf{k.q}}C_q+{\bf{k.(k-q)}}C_{|{\bf{k-q}}|}]\delta \rho_{\bf{q}}(t)\delta \rho_{{\bf{k-q}}}(t).\nonumber\\
\end{eqnarray}

Then we can rewrite Smoluchowski equation (Eq.(\ref{eq7exc})) in Fourier space as,
\begin{eqnarray}
 &\xi&\frac{\partial \delta \rho_{\bf{k}}}{\partial t}=-k_BTk^2(1-\rho_0C_k)\delta \rho_{\bf{k}}(t)\nonumber\\
 &+&\frac{k_BT}{2}\int_q{\bf { dq}}[{\bf{k.q}}C_q+{\bf{k.(k-q)}}C_{|\bf{k-q}|}]\delta \rho_{\bf{q}}(t)\delta \rho_{{\bf{k-q}}}(t).\nonumber\\
\label{mct}
\end{eqnarray}

Eq.(\ref{mct}) can be identified as a Langevin equation for the density field. The last non-linear term on the right hand side is 
 the fluctuating force. In non-Markovian limit, the friction $\xi$ can be replaced by short time part of friction $\gamma$ and 
 a memory function $\mathcal{M}(k,t)$. 
 Thus we can rewrite Eq.(\ref{mct}) as,
  \begin{eqnarray}
 &\gamma& \frac{\partial \delta \rho_{\bf{k}}(t)}{\partial t}+ \frac{k_BTk^2}{S(k)}\delta \rho_{\bf{k}}(t)\nonumber\\
 &+&\int_0^t ds\mathcal{M}(k,t-s)\frac{\partial \delta \rho_{\bf{k}}(t)}{\partial s}+\mathcal{R}_{\bf{k}}(t)=0,
  \label{langevin-1}
 \end{eqnarray}
 where $\mathcal{R}_{\bf{k}}(t)$ is the new thermal noise.
Applying fluctuation dissipation relation (FDR)\cite{kawasaki}, we can write the memory function as,
\begin{eqnarray}
  \mathcal{M}(k,t)=\frac{\langle\mathcal{R}_k(t)\mathcal{R}_{-k}(0)\rangle}{k_BTV}&&\nonumber\\
  =\frac{1}{k_BTV}(\frac{k_BT}{2})^2\frac{1}{(2\pi)^6}\int{\bf { dq}}{\bf{dq'}}\hat k.&[&{\bf{q}}C_q+{\bf{(k-q)}}C_{|\bf{k-q}|}]\nonumber\\
    \times \hat k.[{\bf{q'}}C_{q'}&+&{\bf{(-k-q')}}C_{|\bf{-k-q'}|}]\nonumber\\
   \times \langle\delta \rho_{\bf{q}}(t)\delta \rho_{{\bf{k-q}}}(t)&\delta& \rho_{\bf{q'}}(0)\delta \rho_{{\bf{-k-q'}}}(0)\rangle.\nonumber\\
 \label{FDR-relation}
\end{eqnarray}

 Using the standard method \cite{kawasaki} we can now write a mode coupling theory equation in the over-damped limit for density-density 
 correlation $S_k(t)=\langle\delta \rho_k(t)\delta \rho_{-k}(0)\rangle$ as, 
 \begin{equation}
 \gamma \frac{\partial S_k(t)}{\partial t}+\frac{k_BTk^2}{S_k}S_k(t)+\int \mathcal{M}(t-\tau)\dot S_k(t) d\tau=0.
\label{mct2}
\end{equation}

Where we use Gaussian decoupling and Wick's theorem to treat the four-point 
 correlation function in Eq.(\ref{FDR-relation}).
Note that the numerical
calculations are done for binary systems using well known binary MCT equation \cite{szamel,unravel}.